\documentstyle [aps,multicol,prl,epsf]{revtex}
\begin {document}
\draft

\title {Electronic structure and exchange interactions  of the ladder
vanadates \\ CaV$_2$O$_5$ and MgV$_2$O$_5$}

\author {M. A. Korotin$^{(1)}$,  V. I. Anisimov$^{(1)}$, 
T. Saha-Dasgupta$^{(2)}$, and  I. Dasgupta$^{(2)}$}

\address {$^{(1)}$Institute of Metal Physics, 
Ekaterinburg GSP-170, Russia}

\address {$^{(2)}$Max-Plank-Institut f\"ur Festk\"orperforschung,
D-70569 Stuttgart, Federal Republic of Germany}

\date {sent for publication in J.Phys.:Cond.Mat. 30 July 1999}
\maketitle

\begin {abstract}

We have performed {\it ab-initio} calculations of the electronic structure and
exchange couplings in the layered vanadates CaV$_2$O$_5$ and  MgV$_2$O$_5$.
Based on our results we provide a possible explanation  of the unusual magnetic
properties of these materials, in particular the large difference in the spin
gap between CaV$_2$O$_5$ and MgV$_2$O$_5$.

\end {abstract}
\pacs {}

\begin {multicols}{2}

\section {Introduction}

Spin-1/2 ladder models can describe the magnetic behavior of a variety of
quasi-one-dimensional systems~\cite{dagotto}. Examples include the cuprate
materials SrCu$_2$O$_3$~\cite{SrCu23}, LaCuO$_{2.5}$~\cite{LaCu} and
(Sr,Ca)$_{14}$Cu$_{24}$O$_{41}$~\cite{24-41}.  Spin excitations in the isolated
ladders have a finite energy gap, which makes them prototype spin liquids. This
is of interest in relation to high temperature superconductivity, since upon
doping they become resonating-valence-bond liquids, with a spin excitation gap
and a dominant quasi-long range pairing correlations~\cite{dagotto}.

Another example of the spin-1/2 ladder systems are the layered vanadate
compounds CaV$_2$O$_5$ and MgV$_2$O$_5$. Although CaV$_2$O$_5$ and
MgV$_2$O$_5$  have nearly identical  vanadium oxygen planes however their
magnetic properties are  strikingly different. CaV$_2$O$_5$ has a large spin
gap of about 600~K~\cite{cav2o5}, while the spin gap in  MgV$_2$O$_5$ is very
small only about 20~K~\cite{mgv2o5exp}. In contrast to the planar cuprates,
where a hole in the Cu $x^2-y^2$-orbitals results in a strong antiferromagnetic
exchange coupling for the  180$^\circ$ bonds and a weak ferromagnetic one for
the 90$^\circ$ bonds,  the exchange interactions in these vanadates
can be  more complicated as shown in Fig.~1. Even the signs of the
many  exchange couplings are not obvious in these materials. So one has to
resort to {\it ab-initio} numerical calculations  to get information about the
relative as well as  absolute values of the exchange  couplings in these
systems.  The determination of the exchange couplings is crucial to understand
the markedly different  spin gap behavior in these compounds.

In this paper,  we shall report on the {\it ab-initio} calculation of the 
exchange couplings using the LDA+U method and discover that they  are indeed
different in these two compounds, consistent with their magnetic properties. As
the various exchange couplings  are related to the bare hopping matrix
elements, we shall extract them using a recently developed systematic 
downfolding scheme~\cite{tanusri}.  The advantage of the  downfolding method is
that only the  important orbitals referred to as the active  channels are
retained in the basis and the rest are downfolded thereby  providing a single
or few band tight binding model  capable of reproducing  the details of the LDA
bands close to a prescribed  energy, which is usually the Fermi energy. We
shall use such few band models to extract the various  hopping matrix elements.
This will form a basis  to understand the widely different spin gap behavior
in CaV$_2$O$_5$ and MgV$_2$O$_5$.

The remainder of the paper is organized as follows: In section~II we shall
briefly recapitulate the crystal structure of CaV$_2$O$_5$ and MgV$_2$O$_5$. In
section~III we shall present our LDA+U calculations for the various exchange
couplings and compare them with available experimental results. Section~IV will
be devoted to electronic structure calculations based on  the TB-LMTO
method~\cite{tblmto}, followed by the downfolding method to extract the various
hopping matrix elements, in order to explain the different exchange couplings
in these materials. Finally the conclusions are given in section~V.

\begin{figure}
\centerline {\epsfxsize=3.5cm \epsffile{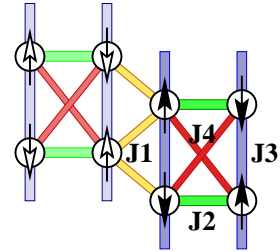} }
\vspace{0.3cm}
\caption {Various magnetic interactions in Ca(Mg)V$_2$O$_5$ compounds.  Two
nearest vanadium ladders with different $z$-coordinates are shown in different
shades.  $J1$, the exchange interaction between nearest V atoms is 
ferromagnetic for CaV$_2$O$_5$ and antiferromagnetic for MgV$_2$O$_5$. $J2$ and
$J3$  are respectively  the antiferromagnetic exchange interactions along the
rung and  leg of the ladder. $J4$ is the antiferromagnetic exchange interaction
between the V atoms along the diagonal of the ladder. The  magnetic structure
used in the calculations is marked with up- and down- arrows. Half of $J1$ and
all of $J4$ exchange interactions are frustrated in this magnetic structure.}
\label{magint} 
\end{figure}

\end{multicols}

\begin{multicols}{2}

\begin {figure} 
\centerline {\epsfxsize=4.2cm \epsffile{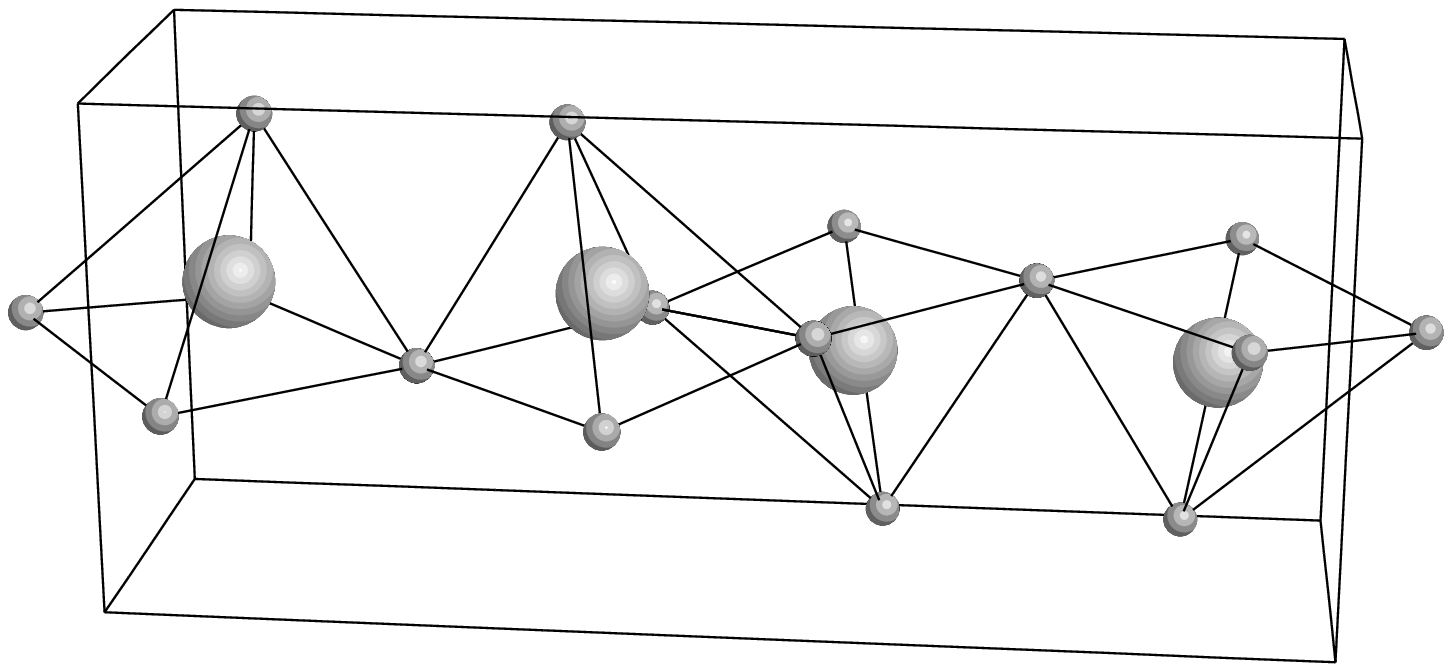} 
             \epsfxsize=4.2cm \epsffile{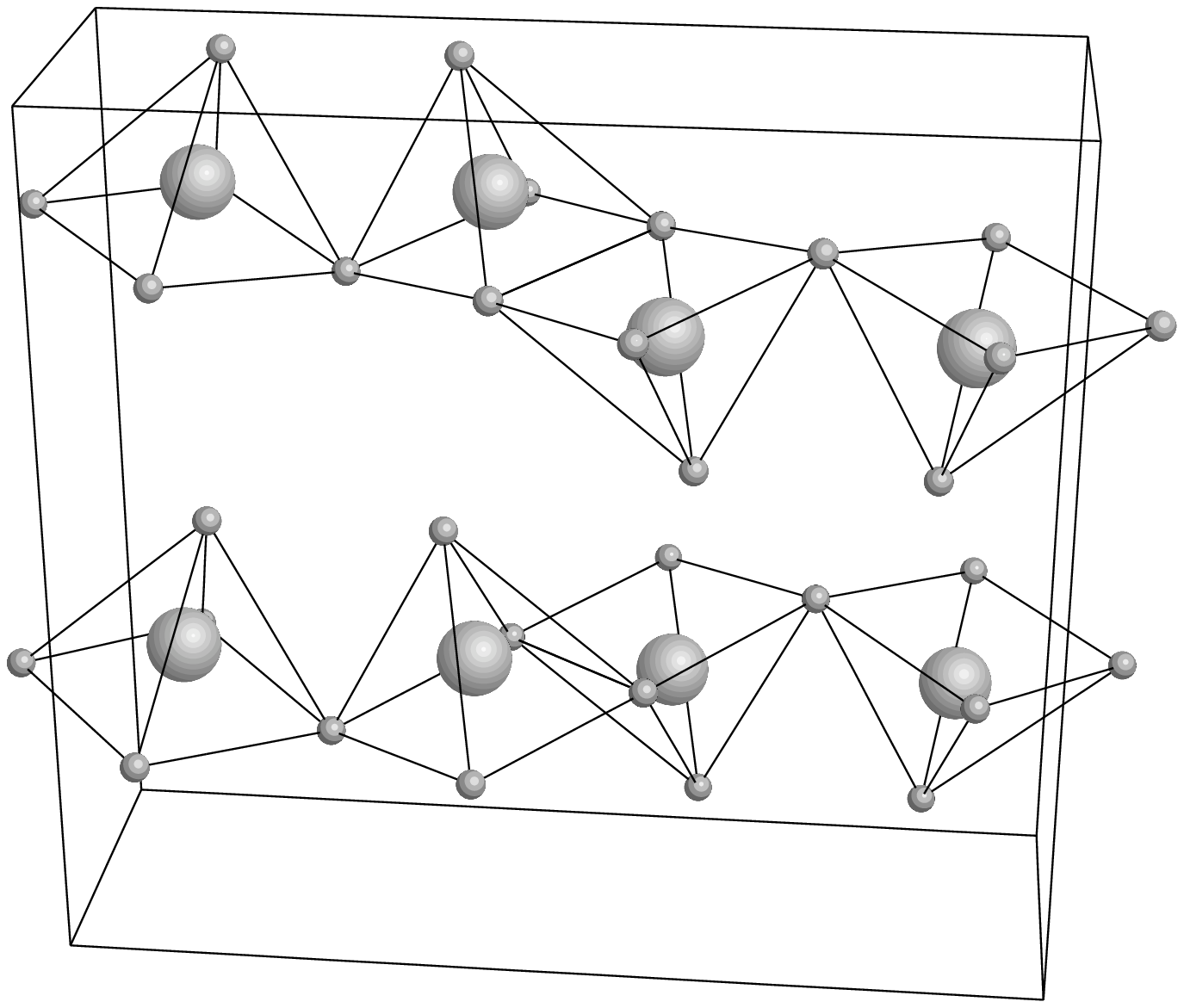} } 
\vspace{0.3cm}
\caption {On the left the crystal structure of CaV$_2$O$_5$ and on the right the
crystal structure of MgV$_2$O$_5$. Ca and Mg atoms are not shown, large balls
represent V and small -- O atoms. In each figure the oxygen atoms constitute a
pyramid which are  linked by edge and corner sharing as described in the text.}
\label{crstr} 
\end{figure}

\section {Structure}

The main building block of the crystal structures of Ca(Mg)V$_2$O$_5$ compounds
are the V ions  roughly in the center of a pyramid of oxygen ions as can be
seen in Fig.~2.  The crystal structure of CaV$_2$O$_5$ is primitive
orthorhombic with space group {\it Pmmn} and lattice constants $a$=11.35~\AA,
$b$=3.60~\AA, and $c$=4.89~\AA. As shown in Fig.~2(left), the structure is
formed by a linkage of VO$_5$ pyramids having apex oxygens in the direction of
the $c$-axis. Oxygen edge- and corner-shared zigzag V chains are formed along
the $b$-axis, where the nearest neighbor V-V distance is 3.03~\AA. Along the
$a$-axis, these chains are linked by sharing corners with the V-V distance of
3.49~\AA. It thus forms a quasi two dimensional ladder layer in the $ab$ plane
with the leg along the zigzag V chains ({\it i.e} along $b$)  while the rung in
the perpendicular direction ({\it i.e.} along $a$). The Ca atoms are located
between the layers and are surrounded by eight O atoms.

The crystal structure of MgV$_2$O$_5$ is base centered orthorhombic with space
group $Cmcm$; and lattice constants $a$=11.02~\AA, $b$=3.69~\AA, and
$c$=9.97~\AA. Again, the structure can be described as a linkage of VO$_5$
pyramids having apex oxygens in the direction of the $c$-axis as can be seen in
Fig.~2(right).  The V zigzag chains extend along the $a$-axis by sharing edges
and corners of the pyramids and  the nearest neighbor V-V distance is 2.98~\AA.
They

\begin {figure} 
\centerline {\epsfxsize=5.0cm \epsffile{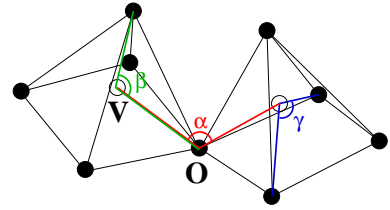} }
\vspace{0.3cm}
\caption {A pair of corner shared VO$_5$ pyramids. The various angles indicated
in the figure  are quoted in Table~1.}
\label{angles} 
\end{figure}

\noindent are also linked by sharing corners  along the $b$-axis, with the V-V
distance of 3.37~\AA, which again leads to a quasi-two dimensional ladder
layers in the $ab$ plane.  However in contrast to CaV$_2$O$_5$, these layers
stack alternately with the distance of ($c/2$) and the Mg atoms are located
between the layers and each are surrounded by six oxygen atoms. As a
consequence, there is a puckering of the V$_2$O$_5$  layers to accommodate Mg
ions in the tetrahedral co-ordination. 

All the structural data that we have discussed above are summarized  in Table~1
and indicated in Fig.~3.  It should be noted that as a consequence of the
puckering  of the V$_2$O$_5$ layers  the tilting angle of the corner shared
pyramids $\alpha$ (see Fig.~3) is appreciably smaller in MgV$_2$O$_5$ in 
comparison to CaV$_2$O$_5$.

\section {The LDA+U and Exchange couplings.}

The LDA+U method was shown to give good results for insulating transition metal
oxides with a partially filled $d$-shell~\cite{review}. The exchange
interaction parameters can be calculated using a procedure based on the Green
function method, developed by Lichtenstein {\it et.
al.}~\cite{lichtexchange,lichtan}.  This method has been successfully applied
to calculate the exchange couplings in KCuF$_3$~\cite{lichtan} and in layered
cuprates~\cite{preprint}.

The LDA+U method~\cite{review,ci1} is the Local Density Approximation (LDA)
modified  by a potential correction restoring a proper description of the
Coulomb interaction between the localized $d$-electrons of transition metal
ions.  This is written in the form of a projection operator:

\end{multicols}

\begin {table} 

\caption {Structural data for CaV$_2$O$_5$ and MgV$_2$O$_5$ compounds. The lattice
constants (in \AA) are $a$ along the rung of the ladder, $b$ -- along the leg,  and
$c$ -- in the vertical direction. The distances (in \AA) between nearest V ions
between ladders, along the leg, the rung and the diagonal are denoted  as d$_{nn}$,
d$_{leg}$, d$_{rung}$ and d$_{diag}$, respectively. The angle $\alpha$ (Figure~3) 
is between V-O-V where oxygen atom is placed between two vanadium atoms forming the
rung. The angle $\beta$ is that between O-V-O where one is rung oxygen and the other
is apical oxygen. Finally the angle $\gamma$  is O-V-O with both the oxygen atoms in
the leg direction. All angle values are listed in degrees.}

\begin{tabular}{lcc}
\multicolumn{1}{l}{characteristic} &
\multicolumn{1}{c}{CaV$_2$O$_5$} &
\multicolumn{1}{c}{MgV$_2$O$_5$} \\ 
\hline 
$a$, $b$, $c$ & 
11.35, 3.60, 4.89 & 
11.02, 3.69, 9.97 \\ 
d$_{nn}$, d$_{rung}$, d$_{leg}$(=$b$), d$_{diag}$ & 
3.03, 3.49, 3.60, 5.02 
& 2.98, 3.57, 3.69, 5.00 \\
$\alpha$, $\beta$, $\gamma$ & 
132.91, 102.94, 135.29 & 
117.57, 109.23, 141.15\\ 
\end{tabular} 

\label{tab-dist} 
\end{table}

\begin{multicols}{2}
\begin{equation}
\widehat{H}=\widehat{H}_{LSDA}+
\sum_{mm^{\prime }}\mid inlm\sigma \rangle
V_{mm^{\prime }}^\sigma \langle inlm^{\prime }\sigma \mid
\label{hamilt}
\end{equation}
\begin{eqnarray}
V_{mm^{\prime }}^\sigma &=&\sum_{\{m\}}
\{U_{m,m^{\prime \prime }
m^{\prime },m^{\prime \prime \prime }}
n_{m^{\prime \prime }
m^{\prime \prime \prime }}^{-\sigma }
+(U_{m,m^{\prime \prime }
m^{\prime },m^{\prime \prime \prime }}\nonumber
 \\
&&-U_{m,m^{\prime \prime }
m^{\prime \prime \prime },
m^{\prime }} )
n_{m^{\prime \prime }m^{\prime \prime
\prime }}^\sigma \}
-U(N-\frac 12)+J(N^{\sigma}-\frac 12)\nonumber
\end{eqnarray}

where $\mid inlm\sigma \rangle $ ($i$ denotes the site, $n$ the main
quantum number, $l$- orbital quantum number, $m$- magnetic number and
$\sigma$- spin index) are d-orbitals of transition metal ions.  
The density matrix is defined by:
\begin{equation}
n_{mm^{\prime }}^\sigma =-\frac 1\pi \int^{E_F}ImG_{inlm,inlm^
{\prime}}^\sigma(E)dE ,
\label{Occ}
\end{equation}

where $G_{inlm,inlm^{^{\prime }}}^\sigma (E)= \langle inlm\sigma \mid
(E-\widehat{H})^{-1}\mid inlm^{^{\prime }}\sigma \rangle $ are the
elements of the Green function matrix, $N^\sigma
=Tr(n_{mm^{\prime}}^\sigma )$ and $N=N^{\uparrow }+N^{\downarrow }.$
$U$ and $J$ are the screened Coulomb and exchange parameters.
The $U_{mm^{\prime}m^{\prime\prime}m^{\prime\prime\prime}}$
is the screened Coulomb interaction among the $nl$ electrons
which can be expressed via integrals over complex spherical harmonics
and $U$, $J$ parameters.  For the Ca(Mg)V$_2$O$_5$ compounds the
values of these parameters were calculated to be $U$=3.6~eV and
$J$=0.88~eV via the so-called "supercell" procedure~\cite{superlsda}
(in the "supercell" calculation only the $xy$-orbitals were considered 
to be localized
so that all other $d$-orbitals could contribute to the screening).
The calculation scheme was realized in the framework of
the Linear Muffin-Tin Orbital (LMTO) method~\cite{lmto} based on the
Stuttgart TBLMTO-47 computer code.

Based on the Green function method, the inter-site exchange couplings can be
derived as the second derivative of the ground state energy with respect to the
magnetic moment rotation angle~\cite{lichtexchange,lichtan}:
\begin{equation}
\label{exchange}
J_{ij}=\sum_{\{m\}}I_{mm^{\prime }}^i
\chi _{mm^{\prime }m^{\prime \prime}
m^{\prime \prime \prime }}^{ij}
I_{m^{\prime \prime }m^{\prime \prime \prime }}^j
\end{equation}

where the spin-dependent potentials $I$ are expressed in terms of the
potentials of Eq.~(\ref{hamilt}),
\begin{equation}  \label{magpot}
I_{mm^{\prime }}^i=V_{mm^{\prime }}^{i\uparrow }-V_{mm^{\prime
}}^{i\downarrow }.
\end{equation}

The effective inter-sublattice susceptibilities are defined in terms of
the LDA+U eigenfunctions $\psi $ as
\begin{equation}
\label{suscep}
\chi _{mm^{\prime }m^{\prime \prime }%
m^{\prime \prime \prime }}^{ij}=\sum_{
{\bf knn}^{\prime }}\frac{n_{n{\bf k\uparrow }}-n_{n^{\prime }{\bf
k\downarrow }}}{\epsilon _{n{\bf k\uparrow }}-\epsilon _{n^{\prime }{\bf
k\downarrow }}}\psi _{n{\bf k\uparrow }}^{ilm^{*}}\psi _{n{\bf k\uparrow }
}^{jlm^{\prime \prime }}\psi _{n^{\prime }{\bf k\downarrow }}^{ilm^{\prime
}}\psi _{n^{\prime }{\bf k\downarrow }}^{jlm^{\prime \prime \prime *}}.\nonumber
\end{equation}

The LDA+U method is the analogue of the Hartree-Fock (mean-field) approximation
for a degenerate Hubbard model~\cite{review}. While in the multi-orbital case
a mean-field approximation gives reasonably good estimates for 

\begin {figure} 
\centerline {\epsfxsize=5.6cm \epsffile{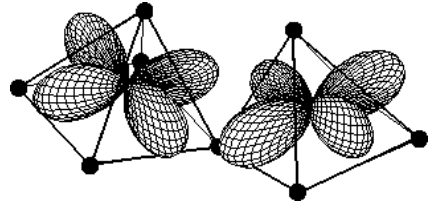} }
\vspace{0.3cm}
\caption {The angular distribution of the V $d$ electron spin density for two V
atoms belonging  to the rung. Oxygen pyramids enclosing the V atoms are also
shown  in the Figure.}
\label{orb} 
\end{figure}

\noindent the total
energy, for the non-degenerate Hubbard model it is known to underestimate the
triplet-singlet energy difference (and thus the value of the effective exchange
coupling $J_{ij}$) by a factor of two for a two-site problem
($E_{HF}=\frac{2t^2}{U}$ and $E_{exact} =\frac{4t^2}{U}$, where $t\ll U$ is
inter-site hopping parameter).

As we discussed in section~II the main building block of the crystal structures
of the Ca(Mg)V$_2$O$_5$ compounds are the  V ions  roughly in the center of a 
pyramid of oxygen ions. The relevant point group symmetry is $C_{4v}$. The
five $d$-orbitals of the vanadium ion transform according to the following
irreducible representations: $3z^2-r^2$ ($A_1$), $x^2-y^2$ ($B_1$), $xy$
($B_2$) and $(xz,yz)$ ($E$). The lowest energy orbital is the V$3d$-orbital of
$xy$-symmetry (using a convention where the axes of the coordinate system is
directed towards the oxygen ions), which is the orbital whose lobes point at
the directions, where the overlap with the oxygen is the smallest as can be
seen in  Fig.~4.

Due to the crystal field splitting, the degeneracy of the V3$d$-shell  is
lifted and the single $d$ electron of V$^{4+}$ ion occupies this $xy$-orbital,
which reminds us of the cuprates, with a single hole in the $x^2-y^2$-orbital. 
The important difference is that while in cuprates all copper atoms are in the
same $(x,y)$-plane as the $x^2-y^2$-orbital, in these vanadates the vertices of
the pyramids point up and down alternately with respect to the basal plane.
Thus the V ions in their centers are correspondingly above and below the
central plane, as can be seen in Fig.~2. As the $xy$-orbitals are parallel to
this plane, so the overlap (and hence the exchange couplings) are expected to
be stronger for vanadium ions situated on the same side of the ladder plane. 
We will show that this is indeed the case. In addition to this alternation, a
tilting of the pyramids is present in the crystal structure of these compounds,
which we shall see seriously influences the interactions.

Another important difference with the cuprates is that 

\begin {table}

\caption {Calculated exchange coupling parameters (in~K).
The "Minus" sign indicates ferromagnetic exchange.}

\begin{tabular}{ccc}
\multicolumn{1}{c}{ } &
\multicolumn{1}{c}{CaV$_2$O$_5$} &
\multicolumn{1}{c}{MgV$_2$O$_5$} \\
\hline
\multicolumn{1}{c}{$J1$}&
\multicolumn{1}{c}{--28} &
\multicolumn{1}{c}{60} \\
\multicolumn{1}{c}{$J2$}&
\multicolumn{1}{c}{608} &
\multicolumn{1}{c}{92} \\
\multicolumn{1}{c}{$J3$}&
\multicolumn{1}{c}{122} &
\multicolumn{1}{c}{144} \\
\multicolumn{1}{c}{$J4$}&
\multicolumn{1}{c}{20} &
\multicolumn{1}{c}{19} \\
\end{tabular}

\label{valJ}
\end{table}

\noindent the $xy$-orbital has a
$\pi$-overlap with the in-plane oxygen atoms in contrast to a much stronger
$\sigma$-overlap in the case of $Cu^{2+}$. Consequently, one can expect much
weaker exchange interaction in vanadates as compared to cuprates. However it is
surprising that the spin gap in CaV$_2$O$_5$ (616~K~\cite{cav2o5}), is {\it
larger} than the typical values for the similar cuprate ladders 
($\approx$460~K~\cite{SrCu23}).

So in the problem under consideration there are two types of contributions to
the exchange interaction parameters $J_{ij}$.  The first one is due to the
$xy-xy$ orbital hopping, and as only this orbital is half-filled this
contribution directly corresponds to the non-degenerate Hubbard model and its
value must be multiplied by a factor of two to correct the Hartree-Fock value.
Other contributions are due to the hoppings to all other orbitals and as the
mean-field approximation is much better for multi-orbital model this part can
be used without modification.

As mentioned earlier, the strongest interaction must be between V atoms which
are situated on the same side of the plane (above or below) [see Fig.~1]. These
atoms form ladders with interactions along the rung and the leg of the ladder
denoted  as $J_2$ and $J_3$ respectively and the interaction between the
ladders as $J_1$ (the notations are chosen to reflect the inter-atomic
distances; the shortest one is between the atoms on different sides of the
plane).

Our calculated values of the exchange couplings are presented in  Table~II. It
can be immediately seen that indeed the strongest interactions are between
atoms on the same side of the plane (the ladder exchanges $J_2$, $J_3$). There
is very strong anisotropy between the  exchange interactions along the rung
($J2$=608~K) and the leg ($J3$=122~K)  for CaV$_2$O$_5$. However for
MgV$_2$O$_5$ the rung ($J2$=92~K) and the leg ($J3$=144~K) exchange interaction
parameters are comparable in size.

Our results suggests CaV$_2$O$_5$ is a system of weakly coupled dimers along
the rung of the ladder with a very strong interaction inside the dimer. The
analysis~\cite{trellis} based on fitting the results of model calculations to
the experimental susceptibility measurements for CaV$_2$O$_5$ confirms the
coupled dimer picture and one of the obtained set of parameters ($J2$=665~K,
$J3$=135~K, $J1$=-25~K) is very close to our {\it ab-initio} calculated
parameters values. However for MgV$_2$O$_5$ our calculations suggest
$\frac{J2}{J1}$=1.53 and $\frac{J3}{J1}$=2.40 which puts MgV$_2$O$_5$ outside
the  scope of the ladder limit,  consistent with the helical ordered gapless
phase  according to the phase diagram obtained by the  Schwinger-boson mean
field theory~\cite{boson}. Recently~\cite{QMC} the exchange parameters for
CaV$_2$O$_5$ and MgV$_2$O$_5$ obtained in the LDA+U method were used for the
calculations of the uniform susceptibility of the Heisenberg model by the
quantum Monte Carlo method. The results agree very well with the experimental
measurements, and particularly very good agreement has been found for
CaV$_2$O$_5$.

\section {LDA band structures and hopping integrals.}

In this section,  we shall investigate the origin of the strong anisotropy of
the  rung and leg exchange interactions in CaV$_2$O$_5$ and its absence  in
MgV$_2$O$_5$ using LDA band structure calculations.  We shall  use a systematic
downfolding scheme to obtain an effective  single (or few) band model
Hamiltonian capable of reproducing  the details of the LDA bands close to the
Fermi level.  We shall extract the  various hopping integrals which in turn
could be related  to the exchange interactions that we have calculated  in the
preceding section.

Fig.~5(a) and Fig.~5(b) shows the energy bands and  density of states (DOS)
respectively for CaV$_2$O$_5$.  The bands are plotted along the various high
symmetry points~\cite{bradley} of the Brillouin zone corresponding to the
primitive orthorhombic lattice.  Similarly  in Fig.~6(a) and Fig.~6(b) we show
the energy bands  and density of states respectively  for MgV$_2$O$_5$. The
bands are now plotted  along the  various directions of the Brillouin zone 
corresponding to base centered orthorhombic lattice. All the energies in the
figures are  measured with respect to the Fermi level of the respective
compounds. In both compounds the bands below -3~eV have predominantly  oxygen
2$p$ character and are separated from the V $d$ complex  by a gap. From -1~eV
to 3~eV the bands with V 3$d$ characters are spread. In Fig.~5(c) and Fig.~6(c)
we show

\begin {figure} 
\centerline {\epsfxsize=8.6cm \epsffile{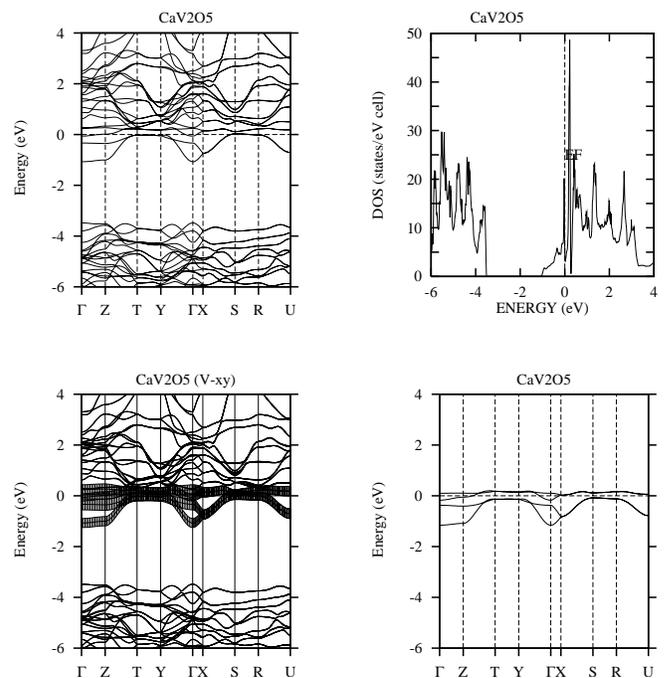} }
\vspace{0.3cm}
\caption {Top left (a) TBLMTO-ASA energy bands; Top right (b) Density of states;
bottom left (c) xy orbital projected band structure and bottom right (d) Band
structure of the effective four band model for CaV$_2$O$_5$.}
\label{bndca} 
\end{figure}

\begin {figure} 
\centerline {\epsfxsize=8.6cm \epsffile{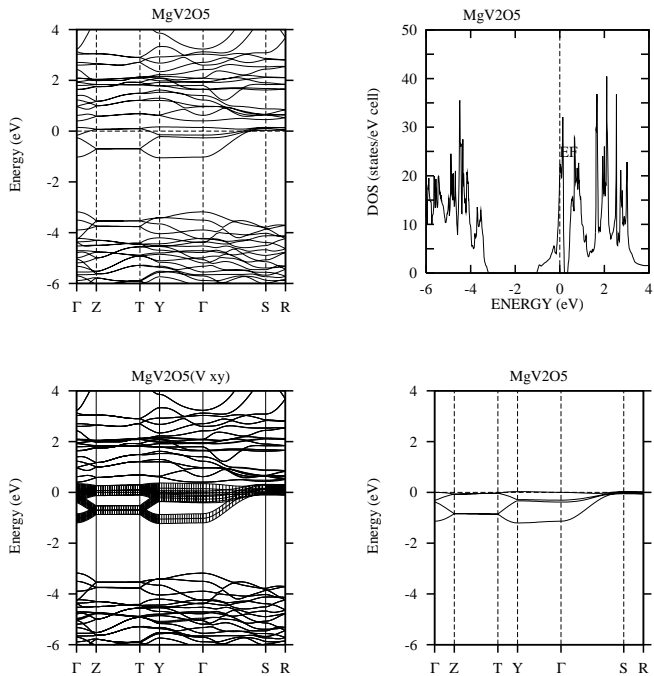} }
\vspace{0.3cm}
\caption {Top left (a) TBLMTO-ASA energy bands; Top right (b) Density
of states; bottom left (c) xy orbital projected band structure and
bottom right (d) Band structure of the effective four band model for
MgV$_2$O$_5$.}
\label{bndmg} 
\end{figure}

\noindent the band structure  of CaV$_2$O$_5$ and MgV$_2$O$_5$ respectively   but
projecting out the V $xy$ orbital character, so that the fatness in each figure
is  proportional to the character of the orbital  V $xy$ (where $x$ runs along
the  rung  and $y$ runs along the leg of the two dimensional ladder system)  in
the wavefunction.  The orbital analysis of the bands or the so called {\it fat
bands}  confirm that the four lowest bands of the V 3$d$ manifold  are
predominantly formed by V $xy$ orbitals,  consistent with the crystal field
arguments presented  in the previous section. The crystal field splitting
between the $xy$-orbital and other $3d$-orbitals is so strong, that in the LDA
band structure, particularly for MgV$_2$O$_5$, the $xy$-bands are separated
from the rest of the V 3$d$-bands by a small energy gap as can be seen in
Fig.~5b. With one subset of energy bands  so well separated  from the rest, 
one can hope that a tight binding model with a single $xy$-orbital  per V site
({\it i.e.} four  band model because there are four V atoms in the unit cell) 
should provide a good approximation to the full band structure  close to the
Fermi level. In order to achieve this, the third generation TB-LMTO 
downfolding method~\cite{tanusri} has been employed to obtain  a four band
effective V-V model. The advantage of the new LMTO method over the previous
ones  is that the resulting  downfolded Hamiltonian is obtained in an 
orthogonal basis and accurately  reproduces the full LDA bands close to the
Fermi level. The energy  bands obtained from the effective four band V-V model
are shown in  Fig.~5(d) and Fig.~6(d) respectively for  CaV$_2$O$_5$ and
MgV$_2$O$_5$. It can be seen that the agreement with the LDA bands in 
Fig.~5(a) and Fig.~6(a) is remarkable. At this point it may be remarked that
the downfolded orbitals  are not thrown away but are included in the tails  of
the active LMTO's which are retained in the basis, {\it i.e.} in the tails of
the V$_{xy}$ orbitals.  In contrast to the fitting procedure  often used to
obtain tight-binding Hamiltonians the present method is deterministic and is
free  from adjustable parameters and also provides the information about  the
wavefunctions. The Fourier transform of the downfolded  Hamiltonian
H(k)$\rightarrow$H(R) gives the effective  hopping parameters for both
compounds. Such an  effective Hamiltonian is long ranged and has been called
the  physical Hamiltonian. We list in Table~3, all the hopping integrals  which
are relevant to understand the various  exchange couplings presented in 
section~III. It can be seen from Table~3, that analogously  to the exchange
couplings the rung and leg hoppings of  CaV$_2$O$_5$ are highly anisotropic
while all the hoppings are  of comparable strength in MgV$_2$O$_5$.

The exchange interaction parameter for the Hubbard model with strong Coulomb
interaction can be estimated as $J=4t^2/U$, where $t$ is the hopping parameter
and $U$ is the  Coulomb interaction. For CaV$_2$O$_5$ the hopping along the
rung of the ladder is 0.252~eV and $U$=3.6~eV, which gives $J$=816~K (our LDA+U
calculation gives $J$=608~K). The ratio of the rung and leg exchange parameters
($J2/J3$) is equal to 4.98 while the ratio of the square of hopping parameters
calculated in the downfolding procedure is equal to 6.25, confirming the strong
anisotropy of the exchange couplings  in CaV$_2$O$_5$. The rung and leg 
hopping parameters for MgV$_2$O$_5$ are nearly equal,  which again agrees with
the LDA+U estimate for the exchange couplings  in MgV$_2$O$_5$.

The hopping integrals extracted from the effective four band  V-V model are
consistent with the exchange couplings calculated  from the LDA+U method in the
previous section. We shall now employ the  downfolding method to explore the
reason for the so different  exchange as well as hopping integrals along the
rung and leg for  CaV$_2$O$_5$ in comparison to MgV$_2$O$_5$ although for both 
compounds the vanadium oxide planes have nearly the same geometry.

It has been argued~\cite{millet} that at least the rung $J2$ and the leg $J3$ 
exchange integrals are mediated by the super-exchange  mechanism through the O
2$p$ orbitals and the size of this exchange integral primarily depends on the 

\begin {table}

\caption {Calculated hopping parameters in an effective four
band V-V model (in~eV). The notations for the hopping parameters 
are same as that for the exchange interactions.}

\begin{tabular}{ccc}
\multicolumn{1}{c}{ } &
\multicolumn{1}{c}{CaV$_2$O$_5$} &
\multicolumn{1}{c}{MgV$_2$O$_5$} \\
\hline
\multicolumn{1}{c}{$t1$}&
\multicolumn{1}{c}{0.076} &
\multicolumn{1}{c}{0.128} \\
\multicolumn{1}{c}{$t2$}&
\multicolumn{1}{c}{0.252} &
\multicolumn{1}{c}{0.114} \\
\multicolumn{1}{c}{$t3$}&
\multicolumn{1}{c}{0.101} &
\multicolumn{1}{c}{0.109} \\
\multicolumn{1}{c}{$t4$}&
\multicolumn{1}{c}{0.056} &
\multicolumn{1}{c}{0.069} \\
\end{tabular}

\label{valt}
\end{table}

\noindent
hopping integral t$_{pd}$ between the $xy$ orbital of the vanadium  and the $p$
orbitals of the oxygen.  According to the canonical band theory  the structural
difference between the two compounds should  account for the observed
differences. It should be noted that  such super-exchange processes are
explicitly taken into account  in our effective V-V model in the process of
downfolding.

Moreover in contrast to the cuprates, where the effective hopping 
predominantly originates from the  $\sigma$-overlap of the Cu3$d$-orbitals with
the oxygen 2$p$ orbitals,  for the vanadates the $\pi$ overlap with the
oxygen  orbitals and the direct V $3d-3d$ overlap could be of the same order
of  magnitude. 

The preceeding discussion suggests a model with oxygen p$_x$ orbital along the
leg and oxygen p$_y$ orbital along the rung in addition to the V $xy$ orbitals 
should be a good starting  point to understand these materials. Such a model
Hamiltonian with more orbitals is usually short ranged and will be referred  to
as a Chemical Hamiltonian, as it is expected to possess the necessary degrees
of freedom so that its tight binding parameters behave in a meaningful way when
the structure is deformed  and when we proceed to study similar materials.
Accordingly, we have extracted all the hoppings in a  tight-binding Hamiltonian
where in the basis we have retained only oxygen p$_x$ orbitals along the leg,
oxygen p$_y$ orbitals along the rung and all the V$_{xy}$ orbitals. All other
orbitals were downfolded. Our calculation  shows V$_{xy}$-O$_p$ hoppings
(t$_{pd}$) are consistent with the prediction of the canonical band theory. In
fact the ratio t$_{pd}^{rung}$  for CaV$_2$O$_5$ to  MgV$_2$O$_5$ is given to 
be 1.21 according to canonical band theory while our downfolding method yields
1.11. Further in this model the anisotropy between the rung and leg hopping is
absent in CaV$_2$O$_5$. However for MgV$_2$O$_5$ the direct V$_{xy}$-V$_{xy}$
hopping along the rung is found to be very small  in comparison to
CaV$_2$O$_5$. This is not consistent with  canonical band theory as the V-V
distances in CaV$_2$O$_5$ are smaller in comparison to MgV$_2$O$_5$ (see
table~1). In order to overcome this problem we tried to include more orbitals
in the basis, as has been  done for the High T$_c$ cuprates~\cite{htsc} and
the  Ladder Cuprate SrCu$_2$O$_3$~\cite{ladder}, however the direct V-V as well
as V-O hoppings remained nearly the same as in the V-O model discussed above.
The reason a chemical Hamiltonian could not be defined  for the vanadates in
the same footing as the cuprates may be attributed  to the complicated geometry
of the vanadates. As a consequence the orbitals are deformed in the process of
downfolding thereby ruling out the validity of the simple canonical band
theory. 

However in this paper we have adopted the following  strategy to  overcome this
problem. The chemical intuition suggests that the anisotropy of the leg and
the rung exchange interactions in CaV$_2$O$_5$ and its absence in 
MgV$_2$O$_5$ may be attributed to the following: (a) the chemical composition
of the compounds particularly the smaller ionic radii of Mg in comparison to
Ca, (b) difference in the crystal structure of these materials.

In order to explore these effects  we have considered three different models
for CaV$_2$O$_5$ and have calculated the exchange interactions by the LDA+U
method  as explained earlier and all the hopping parameters for the effective
four band V$_{xy}$-V$_{xy}$ model.  The first model, referred to  as model~1,
Ca is replaced with Mg in CaV$_2$O$_5$, to examine whether  the chemical
composition plays any role in determining the anisotropy  of the exchange
interactions as well as hoppings in these materials. In order to explore the
role of crystal structure  we have considered the following two models. Model~2
is same as model~1 except now the V-V and V-O distances  are changed so that
they are equivalent to MgV$_2$O$_5$.  Finally in model~3 we have not only
changed the V-V and V-O distances  but also the V-O-V  angles are changed so
that they are the same as in MgV$_2$O$_5$.  The results of our calculation  for
the exchange couplings as well as hoppings for the effective  V-V model are
summarized in Table~4. 

From Table~4 we conclude the following: The calculation of the exchange
coupling  and the hopping parameters in model~1 suggests that the change in
chemical composition, {\it i.e} replacing Ca with Mg do not influence the leg
and rung anisotropy as seen in CaV$_2$O$_5$. Similarly from model~2 we conclude
that  bond lengths do not play any role in deciding  the observed anisotropy
between the leg and rung exchange interactions as well as hoppings in
CaV$_2$O$_5$. However calculations based on model~3 clearly shows as soon as
the V-O-V angles are changed the exchange couplings as well as the  effective
hopping parameters are influenced appreciably. In this case, the rung exchange
coupling and also the bare hopping is even smaller in comparison to the leg. We
obviously recover the values obtained for MgV$_2$O$_5$ as soon as the primitive
orthorhombic stacking is changed to base  centered orthorhombic stacking. These
calculations suggests that the different tilting angle of the  VO$_{5}$
pyramids is the cause for the strikingly different magnetic behavior of the two
vanadates considered here.

\begin {table}

\caption {Calculated exchange couplings [in~K] 
and hopping parameters [in~eV] for the models as
described in the text.}

\begin{tabular}{ccccc}
\multicolumn{1}{c}{System } &
\multicolumn{1}{c}{J2} &
\multicolumn{1}{c}{J3} &
\multicolumn{1}{c}{t2} &
\multicolumn{1}{c}{t3} \\
\hline
\multicolumn{1}{c}{Model1}&
\multicolumn{1}{c}{320} &
\multicolumn{1}{c}{92} &
\multicolumn{1}{c}{0.169} &
\multicolumn{1}{c}{0.065} \\
\multicolumn{1}{c}{Model2}&
\multicolumn{1}{c}{466} &
\multicolumn{1}{c}{57} &
\multicolumn{1}{c}{0.199} &
\multicolumn{1}{c}{0.045}\\
\multicolumn{1}{c}{Model3}&
\multicolumn{1}{c}{24} &
\multicolumn{1}{c}{143} &
\multicolumn{1}{c}{0.053} &
\multicolumn{1}{c}{0.107} \\
\end{tabular}

\label{valmodel}
\end{table}
\end{multicols}

\begin{multicols}{2}

\section {CONCLUSIONS} 

We have used the LDA+U method to compute the  exchange couplings in the layered
vanadate compounds CaV$_2$O$_5$ and MgV$_2$O$_5$.  Our calculation shows that a
strong anisotropy exists between the rung and  leg exchange couplings for
CaV$_2$O$_5$ thus making it a system of weakly coupled dimers  along the rung
with strong  interaction inside the dimer, characterized by a large spin gap.
On the other hand the rung and leg exchange couplings  are found to be of
comparable strength for MgV$_2$O$_5$  making it a small spin gap system. We
have applied the recently developed third-generation  LMTO downfolding method
and the subsequently Fourier transformed  the  downfolded Hamiltonian to
extract the tight-binding hopping parameters between  effective
V$_{xy}$-V$_{xy}$ orbitals  for CaV$_2$O$_5$ and MgV$_2$O$_5$, as well as for
three  different model systems. We conclude, that the stronger {\it tilting} of
the VO$_5$ pyramids in MgV$_2$O$_5$ crystal structure in comparison to
CaV$_2$O$_5$ is the reason that the exchange interactions along rung and leg
are nearly  identical in MgV$_2$O$_5$ while they are anisotropic in 
CaV$_2$O$_5$ leading to the strikingly different magnetic properties of these
materials.

\section*{Acknowledgements}

This work was supported by the Russian Foundation for Basic Research (grants
RFFI-98-02-17275 and RFFI-96-15-96598). ID and TSD  will like to thank
Prof.~O.K.~Andersen and Dr.~O.~Jepsen for useful discussions. MAK and VIA
thanks Max-Plank-Institute for hospitality.

\end {multicols}

\end{document}